\documentclass{INTERSPEECH2023}

\interspeechcameraready 

\usepackage{multirow}
\usepackage{subcaption}
\usepackage[
backend=biber,
style=ieee,
maxbibnames=3,
maxcitenames=3,
doi=false,isbn=false,url=false,eprint=false
]{biblatex} 

\defbibheading{bibliography}[\refname]{}

\title{Speaker-Aware Anti-spoofing}
\name{Xuechen Liu{$^1{}^,{}^2$}, Md Sahidullah{$^3$}, Kong Aik Lee{$^4$}, Tomi Kinnunen{$^1$}}
\address{
  {$^1$}School of Computing, University of Eastern Finland, Joensuu, Finland\\
  {$^2$}Universit\'{e} de Lorraine, CNRS, Inria, LORIA, F-54000, Nancy, France\\
  {$^3$}Institute for Advancing Intelligence, TCG CREST, India \\
  {$^4$}Institute for Infocomm Research, A$^\star$STAR, Singapore
}
\email{xuechen.liu@uef.fi, sahidullahmd@gmail.com, lee\_kong\_aik@i2r.a-star.edu.sg, tkinnu@cs.uef.fi}

\begin{document}

\maketitle

\ninept
 
\begin{abstract}
We address \emph{speaker-aware anti-spoofing}, where prior knowledge of the target speaker is incorporated into a voice spoofing countermeasure (CM). In contrast to the frequently used speaker-independent solutions, we train the CM in a speaker-conditioned way. As a proof of concept, we consider speaker-aware extension to the state-of-the-art AASIST (audio anti-spoofing using integrated spectro-temporal graph attention networks) model. To this end, we consider two alternative strategies to incorporate target speaker information at the frame and utterance levels, respectively. The experimental results on a custom protocol based on ASVspoof 2019 dataset indicate the efficiency of the speaker information via enrollment: we obtain maximum relative improvements of 25.1\% and 11.6\% in equal error rate (EER) and minimum tandem detection cost function (t-DCF) over a speaker-independent baseline, respectively. 
\end{abstract}

\section{Introduction}
\label{sec:intro}

Thanks to recent advances in \emph{neural vocoding} of raw speech waveforms~\cite{vandenOord+2016}, modern \emph{text-to-speech} (TTS) allows the flexible generation of artificial speech that sounds like natural human speech~\cite{jia2018transfer,oord2018parallel,shen2018natural}. Combined with parallel developments in speaker information extraction through \emph{neural speaker embeddings}~\cite{Desplanques2020,snyder2018x} to condition waveform generation~\cite{li2021light,baskar2021eat}, modern TTS allows, in principle, to ``put words into anyone’s mouth'' in the voice of a targeted person. 

Despite numerous useful applications, such flexibility raises obvious concerns. First, in the context of biometric authentication, the possibility for an adversary (attacker) to spoof \emph{automatic speaker verification} (ASV) by miscuing oneself as another individual (target) is well known~\cite{wang2020asvspoof}. Second, the potential negative implications of \emph{deepfakes} --- a combination of `deep learning' and `fake' based on adversarial machine learning~\cite{NIPS2014_5ca3e9b1}--- has recently been called to the attention of researchers~\cite{mirsky2021creation} and the general public~\cite{schick2020deepfakes}. We have already seen alerting examples~\cite{bworld}, even if \emph{speech}-related deepfakes have received less attention compared to image- and video-based deepfakes. Deepfakes used for malicious purposes may damage not only the reputation of the targeted individuals but undermine general trust in audio-visual media and biometric technology. To retain the trust, novel protective means are required. 

On the positive side, the importance of being able to differentiate ``real'' inputs from ``fake'' inputs was proactively recognized early on --- way before the concepts of ``adversarial machine learning'', or ``deepfakes'' were introduced. In particular, the biometrics research community has studied various \emph{anti-spoofing} methods to protect biometric systems for more than two decades~\cite{ratha2001enhancing}. \emph{Presentation attack detection} (PAD) systems~\cite{ISOpresentationAtack}, also known as \emph{countermeasures} (CMs), refer to methods aimed at detecting spoofed inputs. 

In this study, ``CM'' refers to a classifier that takes speech input(s) and produces a binary bonafide/spoof prediction. Since 2015, the ASVspoof challenge series~\cite{wu2017asvspoof} has spearheaded benchmarking of speech CMs using common data and performance metrics \cite{Kinnunen2020-tdcf-fundamentals}. Despite its title, the ASVspoof challenges focus on \emph{standalone}, speaker-independent CMs that can be integrated with ASV systems or other applications. Thanks to the common data provided by the ASVspoof challenge and other similar recent initiatives ~\cite{yi2022add,zhang2021fmfcc}, several standalone speech CMs have been developed ranging from early statistical methods~\cite{todisco2017constant,sahidullah16_interspeech} to recent deep architectures~\cite{wang21fa_interspeech,jung2022aasist, arawnet}. 

\begin{figure*}[t]
    \centering
    \includegraphics[width=0.825\linewidth]{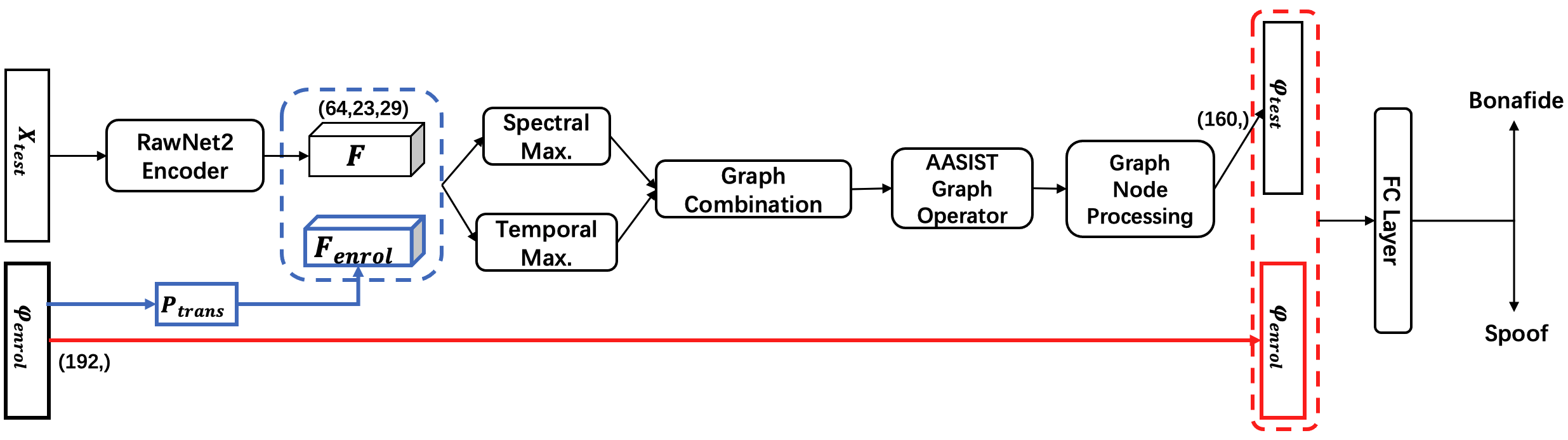}
    \caption{Illustration of speaker information integration via the enrollment vectors. Blue lines and red lines correspond to different approaches which are not applied simultaneously. Dash lines represent the auxiliary attachment operation. Best viewed in color.}
\label{fig:pipeline}
\vspace{-0.25cm}
\end{figure*}

Unfortunately, most existing CMs are far from perfect, particularly when faced with the unknown --- be it unseen vocoders, TTS systems, data domains, or codecs~\cite{Chen2020}. The unconstrained form of the standalone speaker-independent CM task, combined with an artificial speech that is already difficult to differentiate from a real speech by listeners, makes CM generalization beyond training data challenging. The quest for fully general, speaker-independent CM implies that one has to compensate for the potential confounding effects due to speaker, content, and channel variation, with limited prior knowledge. 

Even if not addressed in challenges like ASVspoof, in many applications we \emph{do} have prior knowledge of the target person that could readily be utilized by the CM: spoofing attacks are typically targeted against a particular individual --- the \emph{same} individual whose identity we seek to verify and who the ASV system already `knows' based on enrollment data collected earlier. Concerning deepfakes targeted against public figures such as politicians and news anchors, it seems equally safe to assume that we know \emph{who} the intended target in a potential deepfake sample is. For these reasons, it seems then very reasonable to inform CM at the test time of the identity of the hypothesized speaker based on the enrollment sample. To this end, we present an initial investigation on the use of target speaker information for anti-spoofing that we dub \emph{speaker-aware anti-spoofing}. 

Our study is not the first one to explore this general idea. The two prior studies~\cite{suthokumar2020analysis,castan22_odyssey} that the authors are aware of focus either on replay attack detection with Gaussian mixture model (GMM) backend~\cite{suthokumar2020analysis} or on improving the back-end of the ASV system~\cite{castan22_odyssey}. Our work differs substantially from both studies in terms of CM solutions (statistical model \cite{suthokumar2020analysis} vs. deep learning), the type of fake data (replay attacks \cite{suthokumar2020analysis} vs. synthetic media), the experimental setup, and the evaluation in terms of protocol design and metrics. The main novelty of our work is to propose a precise formulation of the speaker-aware anti-spoofing problem. We also compare different alternative ways of integrating target speaker information into the state-of-the-art AASIST model \cite{jung2022aasist}. To be specific, this information is presented using deep speaker embedding and integrated into different parts of AASIST as illustrated in Fig.~\ref{fig:pipeline} and detailed in the next section.

\begin{figure}[th]
    \centering
    \includegraphics[width=0.7\linewidth]{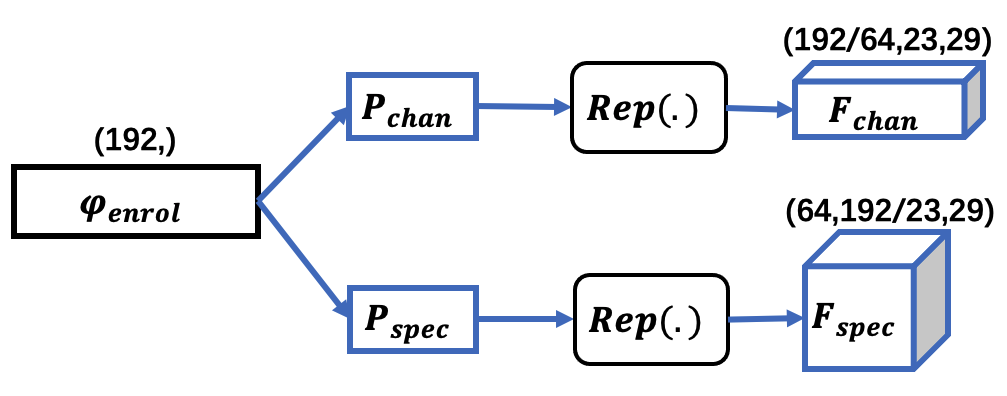}
    \caption{Illustration of transformation of speaker enrollment vector into channel-wise or spectral-wise attachable components. $\mathbf{F}_{\text{enr}}$ in Fig.~\ref{fig:pipeline} can be either $\mathbf{F}_{\text{chan}}$ or $\mathbf{F}_{\text{spec}}$. The transformation matrix $\mathbf{P}_{\text{trans}}$ can be respectively either $\mathbf{P}_{\text{chan}}$ or $\mathbf{P}_{\text{spec}}$. $Rep(.)$ denotes repeating operation on the other two dimensions. Best viewed in color.}
\label{fig:enroll}
\vspace{-0.25cm}
\end{figure}

\section{Speaker-Aware Training of Countermeasure}

\subsection{Problem Definition}
We first define the problem of \emph{speaker-aware anti-spoofing}, as a \textbf{binary classification task of discriminating between bonafide and spoofed speech conditioned on the enrolled speaker}. More specifically, it is a conditional hypothesis test defined as follows:
\begin{itemize}
  \item $H_0$: Test sample is bonafide \underline{and} corresponds to the target speaker.
  \item $H_1$: Test sample is spoofed \underline{and} corresponds to the target speaker. 
\end{itemize}
In practice, we address this task by incorporating additional bonafide utterances of the target speaker, detailed next.
It is worth noticing that the two hypotheses are conditioned on the target speaker, which make them different from the conventional definition of anti-spoofing.

\subsection{Speaker-aware anti-spoofing}
The proposed speaker-aware CM is illustrated in Fig.~\ref{fig:pipeline}. The speaker information can be represented in various ways, from raw audio to well-established deep speaker embeddings. In this study, we focus on the latter. We feed the enrollment audio data into a pre-trained ASV model (here, ECAPA-TDNN \cite{ecapa_tdnn2020}). For each enrollment speaker, we extract speaker embeddings correspondingly and average them to get one enrollment embedding: $\varphi_{\text{enrol}} = \frac{1}{N}\sum_{i=1}^{N}\varphi_{i}$, where $N$ is the number of utterance available for the enrollment. 

We consider the recent AASIST~\cite{aasist2022} for this study. It consists of a speech encoder based on RawNet2~\cite{rawnet2}; two heterogeneous graph attention layers operated respectively on spectral and temporal axes; and a graph pooling layer. The pooling layer is followed by node stacking and a fully connected (FC) layer for binary decision-making. As illustrated by the blue lines in Fig.~\ref{fig:pipeline}, we propose to integrate the enrollment embedding $\varphi_{\text{enrol}}$ into the training by regarding it as auxiliary conditioning information. Methods proposed along with their short-handed forms are presented in the followings.

\textbf{Integration at the encoder output}: Firstly we focus on the output of the encoder, which is a 3-dimensional feature map with channel, spectral, and temporal axes. Let us denote the shape of the map as $(d_{\text{c}}, d_{\text{s}}, d_{\text{t}})$. Inspired by earlier works on channel-wise extension \cite{multichannel_asv2018, multichannel_asr2014, pcen} and speaker adaptation on spectral features \cite{ivector_asr2014}, we extend our embedding vector $\mathbf{F}_{\text{enr}}$ at either channel-level or spectral-level as illustrated in Fig.~\ref{fig:enroll}. The $\mathbf{F}_{\text{enr}}$ in Fig.~\ref{fig:pipeline} can thus be either $\mathbf{F}_{\text{chan}}$ or $\mathbf{F}_{\text{spec}}$. $\mathbf{F}_{\text{chan}}$ is of shape $(d_{\text{embed}}, d_{\text{s}}, d_{\text{t}})$ and $\mathbf{F}_{\text{chan}}$ is of shape $(d_{\text{c}}, d_{\text{embed}}, d_{\text{t}})$. Here, $d_{\text{embed}}$ is the dimension of the enrollment vector. Such two methods on attaching at channel or spectral axis are denoted as \textbf{\emph{enc-chan}} and \textbf{\emph{enc-spec}}, respectively.

\textbf{Integration at the encoder output with dimensionality reduction}: Since the dimensionalities between the embedding vector and the original feature map are respectively different ($d_{\text{embed}}$ vs. $d_{\text{s}}$ or $d_{\text{t}}$), one of them may have potentially more impact to model predictions. Therefore, alternatively, we consider including a transformation matrix $\mathbf{P}_{\text{trans}}$ for dimensionality reduction as shown in Fig.~\ref{fig:pipeline}. $\mathbf{P}_{\text{trans}}$ can be either $\mathbf{P}_{\text{chan}}$ or $\mathbf{P}_{\text{spec}}$, accordingly for channel-level and spectral-level attachment, as illustrated in Fig.~\ref{fig:enroll}. In the case where the enrollment vector is firstly reduced to $d_{\text{c}}$ or $d_{\text{s}}$, $\mathbf{P}_{\text{trans}}$ is initialized with normal distribution and jointly optimized along with other learnable components in AASIST, and with shape of $(d_{\text{embed}}, d_{\text{c}})$ (for $\mathbf{P}_{\text{chan}}$) or $(d_{\text{embed}}, d_{\text{s}})$ ($\mathbf{P}_{\text{spec}}$) respectively; in the case where dimensionality reduction is not carried out, $\mathbf{P}_{\text{trans}}$ is an identity matrix with shape of $(d_{\text{embed}}, d_{\text{embed}})$. 
We denote the resulting feature maps by adding the suffix \emph{-reduced}, so the corresponding methods are \textbf{\emph{enc-chan-reduced}} and \textbf{\emph{enc-spec-reduced}}, respectively.

\textbf{Integration at the FC layer input}: Up to this point, we have described the use of enrollment embedding at the early layers of AASIST. As an alternative strategy, we also consider integration before the fully-connected layer, as illustrated in Fig. ~\ref{fig:pipeline}. The input of the FC layer before the decision-making is a utterance-level 160-dimensional vector, denoted as $\varphi_\text{test}$. It has been extracted in earlier works \cite{sasv_baseline_copied2022} for joint optimization with ASV systems. Here we simply append $\varphi_{\text{enrol}}$ to $\varphi_{\text{test}}$, with the input dimension of the FC layer then being $d_{\text{embed}} + 160$. We denote this mean of attachment as \textbf{\emph{utterance}} when presenting the results in Section \ref{sec:results}.

\begin{figure}[ht]
\centering
\subfloat[Speaker-independent setup\label{fig7:baseset}]{
\includegraphics[width=0.65\linewidth]{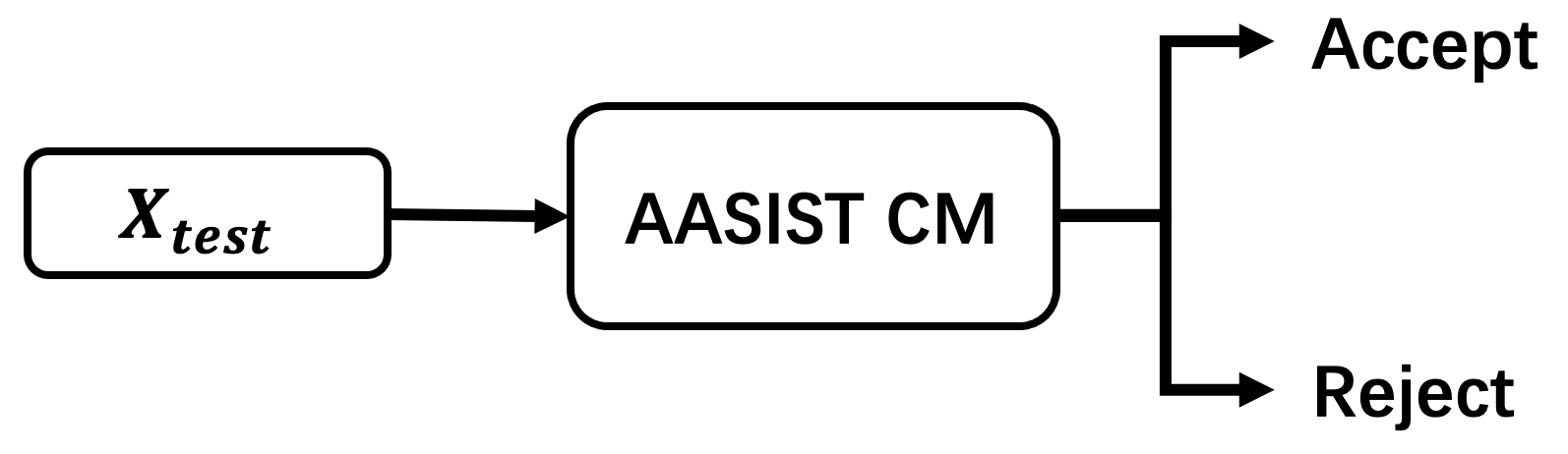} 
} \par
\subfloat[Main \emph{setup}\label{fig7:openset}]{
\includegraphics[width=0.65\linewidth]{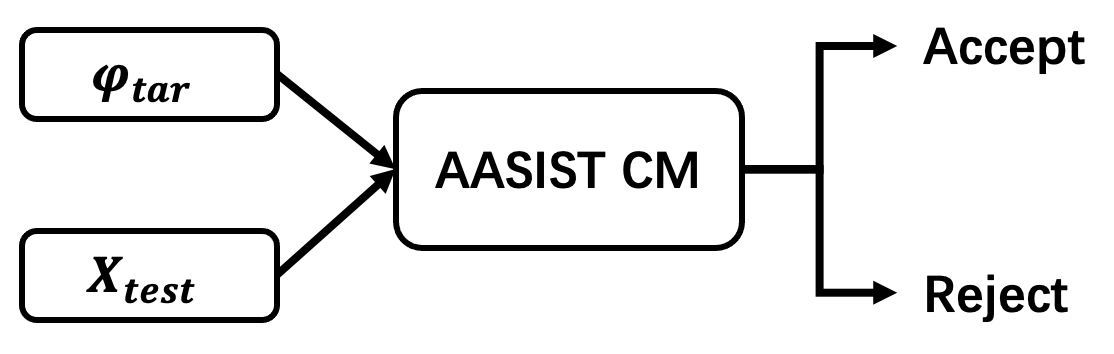} 
} \par
\subfloat[Ablation \emph{setup}\label{fig7:closeset}]{
\includegraphics[width=0.65\linewidth]{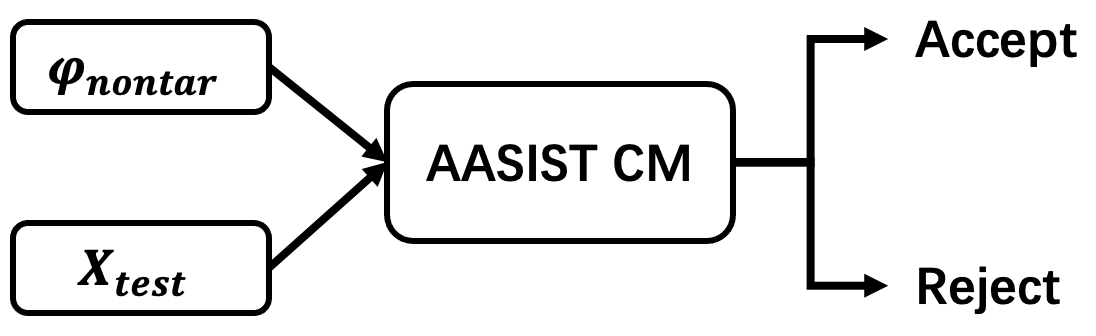} 
}
\caption{The conceptual illustration of the setups for (a) conventional speaker-independent anti-spoofing, (b) speaker-aware anti-spoofing (Main), (c) ablation study setup (Ablation). $\mathbf{\varphi}_\text{tar}$ represents $\mathbf{\varphi}_\text{enrol}$ from the same speaker of the input audio $\mathbf{X}_\text{test}$. $\mathbf{\varphi}_\text{nontar}$ represents $\mathbf{\varphi}_\text{enrol}$ from a different speaker from $\mathbf{X}_\text{test}$. Whereas Main complies with the assumption of known target speaker, Ablation is used to assess the impact of the violation of this assumption on CM performance.}
\label{fig:problem_definitions}
\vspace{-7.5pt}
\end{figure}

\begin{table}[h]
    \centering
    \caption{Number of dev/eval trials available for the full CM and customized protocols.}
      \vspace{-7.5pt}
    \begin{tabular}{|c|c|c|}
    \hline
        Setup & dev & eval \\ \hline
        Original \cite{asvspoof2019} & 24844 & 71237 \\ \hline
        \emph{Main} (Ours) & 23780 & 69252 \\ \hline
    \end{tabular}
    \label{tab:setups}
    \vspace{-7.5pt}
\end{table}

\section{Experimental Setup}
\textbf{Data}. We conduct experiments based on ASVspoof 2019 LA \cite{asvspoof2019}, which covers 19 types of spoofing attacks. The CM training data consists of 20 speakers (9 male, 11 female) and covers 6 attacks (A01-A06), along with the bonafide condition. The CM evaluation data contains other 13 types of attacks (A07-A19). When training the CM model, for each of the 20 speakers, we generate $\mathbf{\varphi}_\text{enrol}$ by averaging over his/her speaker embeddings over $N$ randomly selected utterances from the bonafide condition. We set $N=11$ for female, and $N=19$ for male. 

\textbf{Protocol}. The design logic of the protocol is shown in Fig.~\ref{fig7:openset}, with comparison to the regular anti-spoofing setup shown in Fig.~\ref{fig7:baseset}. We recall our assumption that the test utterance originates from a known target speaker, and the task is to determine whether or not the sample is bonafide or spoofed utterance. Therefore, for each test utterance, its associated speaker embedding for enrollment $\mathrm{\varphi}_\text{enrol}$ is from the same target speaker.
In this case, our evaluation protocol is based on the original ASVspoof 2019 CM protocol (``Original" in Table \ref{tab:setups}), with the bonafide speech trials without the corresponding target speaker in the dataset being removed. We use the speaker information available in the ASV protocol files in the original metadata.
The protocol statistics are presented in Table \ref{tab:setups}. We refer to this protocol setup as \emph{Main}, which differs from the \emph{Ablation} setup described in Section~\ref{secsec:ablation}. 

\textbf{Model.} For the AASIST model, we adopt the solution from the open-sourced repository as the speaker-independent baseline\footnote{\url{https://github.com/clovaai/aasist}}. Model training and hyperparameter setups followed the ones described in \cite{aasist2022}, except for the batch size that was reduced from 24 down to 12 due to limited computational resources (via a single NVIDIA GeForce GTX 2080Ti). The original shape of the feature map was $(d_{\text{c}}, d_{\text{s}}, d_{\text{t}}) = (64, 23, 29)$, as shown in Fig.~\ref{fig:pipeline}. We used the open-sourced pre-trained ECAPA-TDNN\footnote{\url{https://github.com/TaoRuijie/ECAPA-TDNN/}} \cite{ecapa_tdnn2020} as the pre-trained ASV model, to extract the speaker embedding with $d_{\text{embed}} = 192$. Each speaker embedding was extracted from the first fully-connected layer after the pooling layer for each input sentence.

\textbf{Evaluation.} We report \emph{equal error rate} (EER) and minimum \emph{tandem detection cost function} (tDCF) \cite{Kinnunen2020-tdcf-fundamentals}. Since compared to minimum tDCF, EER reflects more on the sole CM performance \cite{aasist2022, Kinnunen2020-tdcf-fundamentals}, we present our analysis on results primarily on EERs, including the per-attack-type analysis.

\begin{table}[t]
    \centering
    \caption{Results in terms of \emph{pooled} EER and minimum tDCF.}
    \vspace{-7.5pt}
    \begin{tabular}{|c|cc|cc|}
    \hline
        & \multicolumn{2}{|c|}{\emph{Main}} & \multicolumn{2}{|c|}{\emph{Ablation}} \\ \hline
        Method & EER(\%) & tDCF & EER(\%) & tDCF \\ \hline
        (Baseline) & 1.51 & 0.043 & 1.51 & 0.043 \\ \hline
        \emph{enc-chan} & 1.48 & 0.049 & 1.80 & 0.061 \\ \cline{2-5}
        \emph{enc-chan-reduced} & 2.27 & 0.077 & 2.20 & 0.073  \\ \hline
        \emph{enc-spec} & \textbf{1.13} & \textbf{0.038} & \textbf{1.47} & \textbf{0.049} \\ \cline{2-5}
        \emph{enc-spec-reduced} & 1.65 & 0.055 & 1.88 & 0.061 \\ \hline
        \emph{utterance} & 1.89 & 0.059 & 1.78 & 0.052 \\ \hline
    \end{tabular}
    \label{tab:fullset_results}
    \vspace{-0.25cm}
\end{table}

\begin{table*}[t]
    \centering
    \caption{Results in terms of per-attack-type EER(\%) for baselines and best-performed systems. Spoofing attacks in bold font indicate the acquisition of speaker information during the development, according to \cite{asvspoof2019}. The ASV EER is returned by the same pre-trained ECAPA-TDNN model as used in this study, as described in \cite{sasv_baseline_copied2022}.}
    \vspace{-7.5pt}
    \begin{tabular}{|c|c|ccccccccccccc|}
    \hline
     & Method & A07 & A08 & A09 & A10 & A11 & A12 & A13 & A14 & A15 & A16 & A17 & A18 & A19 \\ \hline \hline
    
    \multirow{2}{2em}{\emph{Main}} & (Baseline) & 0.75 & 0.19 & 0.02 & 0.88 & 0.37 & 0.72 & 0.14 & 0.15 & 0.47 & 0.73 & 2.15 & 4.80 & 0.78 \\ \cline{2-15}
    & \emph{enc-spec} & 1.18 & \textbf{0.07} & \textbf{0.00} & 1.38 & 0.41 & 0.98 & 0.22 & 0.28 & 0.98 & \textbf{0.65} & \textbf{1.28} & \textbf{2.70} & \textbf{0.34} \\ \hline    \hline
    
    \emph{Ablation} & \emph{enc-spec} & 1.57 & \textbf{0.08} & \textbf{0.00} & 1.95 & 0.47 & 1.26 & 0.28 & 0.35 & 1.30 & \textbf{0.71} & \textbf{1.79} & \textbf{3.13} & \textbf{0.30} \\ \hline \hline
    
    \multicolumn{2}{|c|}{ASV EER \cite{sasv_baseline_copied2022}} & 32.66 & 18.80 & 2.20 & 50.61 & 47.08 & 39.56 & 11.62 & 35.39 & 36.54 & 60.71 & 1.85 & 2.38 & 4.77 \\
    \hline
    \end{tabular}
    \label{tab:per_cond_results}
    \vspace{-7.5pt}
\end{table*}

\section{Results and Analysis}
\label{sec:results}

\subsection{Results}
Results in terms of pooled EER and tDCF are presented in Table~\ref{tab:fullset_results}. While channel-wise speaker integration without dimensionality reduction only marginally improves the EER, the spectral-wise integration works nicely by achieving the lowest numbers in both metrics, outperforming the baseline by relatively 25.1\% and 11.6\% in terms of EER and minimum tDCF, respectively. This indicates the efficiency of integrating the target speaker integration method on the spectral feature map. The relatively under-performed channel-wise integration, in turn, might be explained by noting the original audio is single channel and contain a rather low level of noise. Applying dimensionality reduction degrades the CM performance for both methods. Attaching the enrollment vector before the FC layer with the bottleneck embedding does not lead to improvements. 

A detailed breakdown of the results per attack is shown in Table~\ref{tab:per_cond_results} for the baseline and the best speaker-aware anti-spoofing approach (per protocol). In addition to the CM results shown in the first four lines, the table also displays the EERs of the ASV system on the full CM protocol. These numbers serve to indicate the effectiveness of each attack in spoofing the ASV system (but should not be compared with the CM results). Reflected by the ASV EERs, some attacks do not spoof the ASV model well, such as A09, A17, and A18, which means that those algorithms do not model the speaker information well. 

Moving back to the CM performances, there are five types of attacks (A08, A09, A16, A17, A18) where the best-performed proposed method outperforms (or reaches similar performance with) the baseline under both protocols. The ASV EER for four of them is relatively low (lower than 20\%) except for A16, which indicates that the proposed speaker information integration method further exploits the weakness of the spoofing algorithm by not being able to encode the target speakers well. Improvements can also be observed on A16, which corresponds to the highest ASV EER among all spoofing algorithms. This may exploit the compensation ability of the proposed algorithms on strong attacks that models speaker information well. Future work may further exploit the relationship between the speaker information modeling ability of the spoofing algorithms and its compensation from the CM via such integration.

\subsection{Ablation study: Mis-specified speaker identity}
\label{secsec:ablation}
The evaluation setup and results described above are based on the assumption that the input audio is target speaker. A natural question that arises is \emph{what might happen if this assumption is violated?} -- i.e. how robust the CM is to modeling misspecification in terms of mismatched speaker identities across the enrollment and test utterances. To this end, in this ablation study, we assume that the bonafide input audio is not from its corresponding speaker. In this case, we retain the exact test utterances as in the main protocol but replace the corresponding enrollment utterance with a randomly selected enrollment utterance from another randomly selected speaker. This ablation setup is illustrated in Fig.~\ref{fig7:closeset}.

The overall results for the proposed methods for this setup are shown in Table \ref{tab:fullset_results}. For most proposed methods, compared to the \emph{Main} setup, the results in both metrics are degraded, but not by a large margin. The EER of the best-performed \emph{enc-spec} degrades by relatively 25.1\%, but still retains the accuracy of the speaker-independent baseline, even in the severe modeling mis-specification / strong violation of modeling assumption. For \emph{enc-chan-reduced} and \emph{utterance}, the results remain at about same level. The per-attack results for \emph{enc-spec} under this setup is shown in Table \ref{tab:per_cond_results}. For the six types of attacks where improvements are observed under the \emph{Main} setup, \emph{enc-spec} holds its superiority over the baseline, although with marginal performance degradation from \emph{Main} except on A09 and A19. While such degradation indicates the usefulness of target speaker information compared to the one from another speaker, the potential of such \emph{non-target} speaker information still deserves further investigation and extension onto other scenarios.

\begin{figure}[t]
    \centering
    \includegraphics[width=\linewidth]{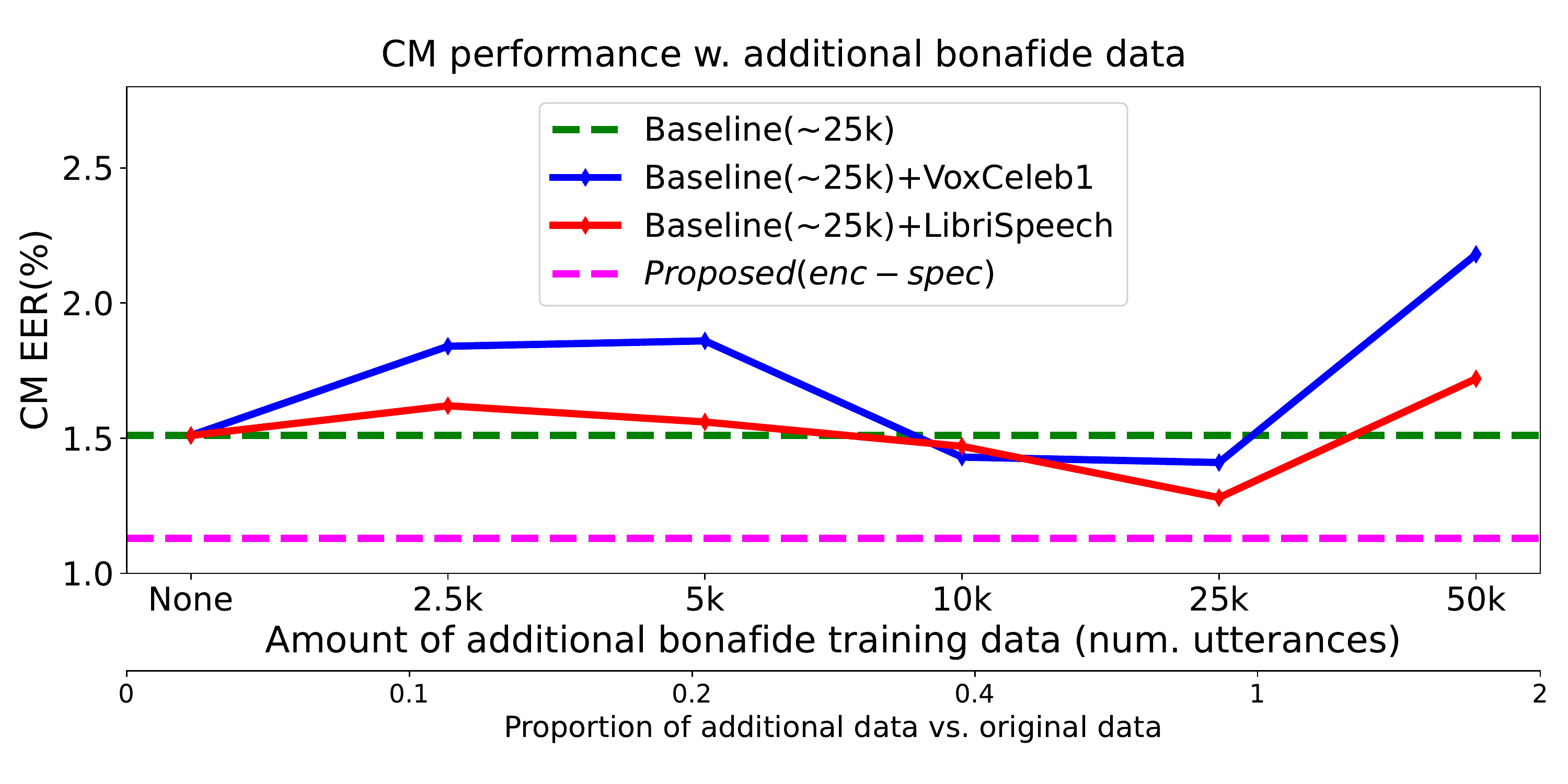}
    \caption{The relationship between the additional data from different common speech processing datasets and the CM performance under Main. The green dashed line indicates the baseline performance and the pink one indicates the best-performed system. Best viewed in color.}
\label{fig:cm_performance}
\vspace{-7.5pt}
\end{figure}

\subsection{Ablation study: Additional bonafide training data}
\label{sec:ablation}
An enrollment vector is not only a speaker representation but also an additional container of bonafide information. Both speaker and bonafide information can be useful as prior conditions for training CM systems. Therefore, we consider an experiment on the effect of additional bonafide training data.

We implement the addition under the full CM protocol by pooling additional speech data from various datasets. We consider VoxCeleb \cite{voxsrc2019} and LibriSpeech \cite{librispeech} corpora. For each dataset, we vary the number of utterances for CM training. The results are shown in Fig.~\ref{fig:cm_performance}, along with the baseline and the best-performing speaker-aware CM. The figure reveals two interesting patterns. First, the larger amount of additional sole bonafide data from either VoxCeleb1 or LibriSpeech improves performance. Second, the baseline performance is improved with 25k additional utterances, where the amount of data added is almost equal to the total amount of CM training data (25380 utterances \cite{asvspoof2019}). However, adding more data does not necessarily lead to better performance. Relating to the results above, this suggests a more significant benefit provided by additional speaker information, but this might also since the ASVspoof dataset is originated from \emph{VCTK}\footnote{\url{https://datashare.is.ed.ac.uk/handle/10283/2651}}, which is a very clean dataset recorded using the anechoic room. Future work may investigate this issue.

\section{Conclusion}
We have investigated the feasibility of \textbf{speaker-aware anti-spoofing} using state-of-the-art AASIST countermeasure for synthetic spoofing attack detection. Our findings indicate that integration of target speaker enrollment embedding as auxiliary information leads to up to 25.1\% relative improvement in anti-spoofing EER. Additional experiments on the effect of alternative speaker information and augmenting the bonafide training using auxiliary corpora have suggested that the proposed speaker-aware training strategy can be more effective. Confirming similar findings done in the two earlier studies using completely different classifiers and datasets \cite{suthokumar2020analysis, castan22_odyssey}, this study adds evidence to the positive impact of target speaker prior information. Future work may focus on the Siamese network to encode speaker information and make it available during the training of the CM module, along with more advanced cohort models to encode the speaker information. 

\section{References}
{
\printbibliography

@INPROCEEDINGS{arawnet,
  author={Teng, Zhongwei and Fu, Quchen and White, Jules and Powell, Maria E. and Schmidt, Douglas C.},
  booktitle={International Conference on Pattern Recognition (ICPR)}, 
  title={{ARawNet}: A Lightweight Solution for Leveraging Raw Waveforms in Spoof Speech Detection}, 
  year={2022},
  volume={},
  number={},
  pages={692-698},
  doi={10.1109/ICPR56361.2022.9956138}}

@article{wang2020asvspoof,
  title={{ASVspoof} 2019: A large-scale public database of synthesized, converted and replayed speech},
  author={Wang, Xin and Yamagishi, Junichi and Todisco, Massimiliano and Delgado, H{\'e}ctor and Nautsch, Andreas and Evans, Nicholas and Sahidullah, Md and Vestman, Ville and Kinnunen, Tomi and Lee, Kong Aik and others},
  journal={Computer Speech \& Language},
  volume={64},
  pages={101114},
  year={2020},
  publisher={Elsevier}
}

@article{sasv_baseline_copied2022,
%       title={A New Fusion Strategy for Spoofing Aware Speaker Verification}, 
%       author={You Zhang and Ge Zhu and Zhiyao Duan},
%       year={2022},
%       eprint={2202.05253},
%       archivePrefix={arXiv},
%       primaryClass={eess.AS}, 
%       journal   = {CoRR},
%       volume    = {abs/2202.05253},
%       url       = {http://arxiv.org/abs/2202.05253},
% }

@article{suthokumar2020analysis,
  title={An analysis of speaker dependent models in replay detection},
  author={Suthokumar, Gajan and Sriskandaraja, Kaavya and Sethu, Vidhyasaharan and Ambikairajah, Eliathamby and Li, Haizhou},
  journal={APSIPA Transactions on Signal and Information Processing},
  volume={9},
  year={2020},
  publisher={Cambridge University Press}
}

@inproceedings{castan22_odyssey,
  author={Diego Castan and Md Hafizur Rahman and Sarah Bakst and Chris Cobo-Kroenke and Mitchell McLaren and Martin Graciarena and Aaron Lawson},
  title={{Speaker-Targeted Synthetic Speech Detection}},
  year=2022,
  booktitle={Proc. The Speaker and Language Recognition Workshop (Odyssey 2022)},
  pages={62--69},
  doi={10.21437/Odyssey.2022-9}
}

@inproceedings{Chen2020,
  author={Tianxiang Chen and Avrosh Kumar and Parav Nagarsheth and Ganesh Sivaraman and Elie Khoury},
  title={{Generalization of Audio Deepfake Detection}},
  year=2020,
  booktitle={Proc. Odyssey 2020 The Speaker and Language Recognition Workshop},
  pages={132--137}
}

@inproceedings{wang21fa_interspeech,
  author={Xin Wang and Junichi Yamagishi},
  title={{A Comparative Study on Recent Neural Spoofing Countermeasures for Synthetic Speech Detection}},
  year=2021,
  booktitle={Proc. Interspeech},
  pages={4259--4263}
}

@inproceedings{jung2022aasist,
  title={{AASIST}: Audio anti-spoofing using integrated spectro-temporal graph attention networks},
  author={Jung, Jee-weon and Heo, Hee-Soo and Tak, Hemlata and Shim, Hye-jin and Chung, Joon Son and Lee, Bong-Jin and Yu, Ha-Jin and Evans, Nicholas},
  booktitle={Proc. ICASSP},
  pages={6367--6371},
  year={2022},
  organization={IEEE}
}

@article{todisco2017constant,
  title={Constant Q cepstral coefficients: A spoofing countermeasure for automatic speaker verification},
  author={Todisco, Massimiliano and Delgado, H{\'e}ctor and Evans, Nicholas},
  journal={Computer Speech \& Language},
  volume={45},
  pages={516--535},
  year={2017},
  publisher={Elsevier}
}

@inproceedings{zhang2021fmfcc,
  title={{FMFCC-a}: a challenging Mandarin dataset for synthetic speech detection},
  author={Zhang, Zhenyu and Gu, Yewei and Yi, Xiaowei and Zhao, Xianfeng},
  booktitle={International Workshop on Digital Watermarking},
  pages={117--131},
  year={2021},
  organization={Springer}
}

@inproceedings{yi2022add,
  title={{ADD} 2022: the first audio deep synthesis detection challenge},
  author={Yi, Jiangyan and Fu, Ruibo and Tao, Jianhua and Nie, Shuai and Ma, Haoxin and Wang, Chenglong and Wang, Tao and Tian, Zhengkun and Bai, Ye and Fan, Cunhang and others},
  booktitle={Proc. ICASSP},
  pages={9216--9220},
  year={2022},
  organization={IEEE}
}

@article{wu2017asvspoof,
  title={{ASVspoof}: the automatic speaker verification spoofing and countermeasures challenge},
  author={Wu, Zhizheng and Yamagishi, Junichi and Kinnunen, Tomi and Hanil{\c{c}}i, Cemal and Sahidullah, Mohammed and Sizov, Aleksandr and Evans, Nicholas and Todisco, Massimiliano and Delgado, Hector},
  journal={IEEE Journal of Selected Topics in Signal Processing},
  volume={11},
  number={4},
  pages={588--604},
  year={2017},
  publisher={IEEE}
}

@techreport{ISOpresentationAtack,
  type = {{Standard}},
  publisher = {{ISO, Geneva, Switzerland}},
  key = {{ISO/IEC 30107}},
  year = {2016},
  title = {{ISO/IEC 30107. Information technology -- biometric presentation attack detection}}
  }

@article{ratha2001enhancing,
  title={Enhancing security and privacy in biometrics-based authentication systems},
  author={Ratha, Nalini K. and Connell, Jonathan H. and Bolle, Ruud M.},
  journal={IBM systems Journal},
  volume={40},
  number={3},
  pages={614--634},
  year={2001},
  publisher={IBM}
}

@misc{bworld,
  title = {{Top 5 Deepfake Scams That Stormed the Internet This Year}},
  howpublished = "\url{https://www.analyticsinsight.net/top-5-deepfake-scams-that-stormed-the-internet-this-year/}",
  year = {2022}, 
  note = "[Online; accessed 10-October-2022]"
}

@book{schick2020deepfakes,
  title={Deepfakes: The coming infocalypse},
  author={Schick, Nina},
  year={2020},
  publisher={Hachette UK}
}

@article{mirsky2021creation,
  title={The creation and detection of deepfakes: A survey},
  author={Mirsky, Yisroel and Lee, Wenke},
  journal={ACM Computing Surveys (CSUR)},
  volume={54},
  number={1},
  pages={1--41},
  year={2021},
  publisher={ACM New York, NY, USA}
}

@inproceedings{NIPS2014_5ca3e9b1,
 author = {Goodfellow, Ian and Pouget-Abadie, Jean and Mirza, Mehdi and Xu, Bing and Warde-Farley, David and Ozair, Sherjil and Courville, Aaron and Bengio, Yoshua},
 booktitle = {Advances in Neural Information Processing Systems},
 editor = {Z. Ghahramani and M. Welling and C. Cortes and N. Lawrence and K.Q. Weinberger},
 pages = {},
 publisher = {Curran Associates, Inc.},
 title = {Generative Adversarial Nets},
 volume = {27},
 year = {2014}
}

@inproceedings{baskar2021eat,
  title={EAT: Enhanced ASR-TTS for self-supervised speech recognition},
  author={Baskar, Murali Karthick and Burget, Luk{\'a}{\v{s}} and Watanabe, Shinji and Astudillo, Ramon Fernandez and others},
  booktitle={Proc. ICASSP},
  pages={6753--6757},
  year={2021},
  organization={IEEE}
}

@inproceedings{li2021light,
  title={Light-TTS: Lightweight Multi-Speaker Multi-Lingual Text-to-Speech},
  author={Li, Song and Ouyang, Beibei and Li, Lin and Hong, Qingyang},
  booktitle={Proc. ICASSP},
  pages={8383--8387},
  year={2021},
  organization={IEEE}
}

@inproceedings{snyder2018x,
  title={X-vectors: Robust dnn embeddings for speaker recognition},
  author={Snyder, David and Garcia-Romero, Daniel and Sell, Gregory and Povey, Daniel and Khudanpur, Sanjeev},
  booktitle={Proc. ICASSP},
  pages={5329--5333},
  year={2018},
  organization={IEEE}
}

@inproceedings{Desplanques2020,
  author={Brecht Desplanques and Jenthe Thienpondt and Kris Demuynck},
  title={{ECAPA-TDNN: Emphasized Channel Attention, Propagation and Aggregation in TDNN Based Speaker Verification}},
  year=2020,
  booktitle={Proc. Interspeech},
  pages={3830--3834}
}

@inproceedings{shen2018natural,
  title={Natural tts synthesis by conditioning wavenet on mel spectrogram predictions},
  author={Shen, Jonathan and Pang, Ruoming and Weiss, Ron J and Schuster, Mike and Jaitly, Navdeep and Yang, Zongheng and Chen, Zhifeng and Zhang, Yu and Wang, Yuxuan and Skerrv-Ryan, Rj and others},
  booktitle={Proc. ICASSP},
  pages={4779--4783},
  year={2018},
  organization={IEEE}
}

@inproceedings{oord2018parallel,
  title={Parallel wavenet: Fast high-fidelity speech synthesis},
  author={Oord, Aaron and Li, Yazhe and Babuschkin, Igor and Simonyan, Karen and Vinyals, Oriol and Kavukcuoglu, Koray and Driessche, George and Lockhart, Edward and Cobo, Luis and Stimberg, Florian and others},
  booktitle={International conference on machine learning},
  pages={3918--3926},
  year={2018},
  organization={PMLR}
}

@article{jia2018transfer,
  title={Transfer learning from speaker verification to multispeaker text-to-speech synthesis},
  author={Jia, Ye and Zhang, Yu and Weiss, Ron and Wang, Quan and Shen, Jonathan and Ren, Fei and Nguyen, Patrick and Pang, Ruoming and Lopez Moreno, Ignacio and Wu, Yonghui and others},
  journal={Advances in neural information processing systems},
  volume={31},
  year={2018}
}

@inproceedings{vandenOord+2016,
author={Aäron van den Oord and Sander Dieleman and Heiga Zen and Karen Simonyan and Oriol Vinyals and Alex Graves and Nal Kalchbrenner and Andrew Senior and Koray Kavukcuoglu},
title={WaveNet: A Generative Model for Raw Audio},
year=2016,
booktitle={9th ISCA Speech Synthesis Workshop},
pages={125--125}
}

@ARTICLE{asvspoof2019,
  author={A. {Nautsch} and X. {Wang} and N. {Evans} and T. H. {Kinnunen} and V. {Vestman} and M. {Todisco} and H. {Delgado} and M. {Sahidullah} and J. {Yamagishi} and K. A. {Lee}},
  journal={IEEE Transactions on Biometrics, Behavior, and Identity Science}, 
  title={ASVspoof 2019: Spoofing Countermeasures for the Detection of Synthesized, Converted and Replayed Speech}, 
  year={2021},
  volume={3},
  number={2},
  pages={252-265},
  doi={10.1109/TBIOM.2021.3059479}
}

@ARTICLE{Kinnunen2020-tdcf-fundamentals,
  author={Kinnunen, Tomi and Delgado, Héctor and Evans, Nicholas and Lee, Kong Aik and Vestman, Ville and Nautsch, Andreas and Todisco, Massimiliano and Wang, Xin and Sahidullah, Md and Yamagishi, Junichi and Reynolds, Douglas A.},
  journal={IEEE/ACM Transactions on Audio, Speech, and Language Processing}, 
  title={Tandem Assessment of Spoofing Countermeasures and Automatic Speaker Verification: Fundamentals}, 
  year={2020},
  volume={28},
  number={},
  pages={2195-2210},
  doi={10.1109/TASLP.2020.3009494}}

@inproceedings{voxsrc2019,
  author={J. Chung and
               A. Nagrani and
               E. Coto and
               W. Xie and
               M. McLaren and
               D. A. Reynolds and
               A. Zisserman},
  title={Vox{SRC} 2019: The first {VoxCeleb} Speaker Recognition Challenge},
  year=2019,
  booktitle={ISCA archive},
}

@inproceedings{multichannel_asv2018,
  author={Danwei Cai and Xiaoyi Qin and Ming Li},
  title={{Multi-Channel Training for End-to-End Speaker Recognition Under Reverberant and Noisy Environment}},
  year=2019,
  booktitle={Proc. Interspeech},
  pages={4365--4369},
  doi={10.21437/Interspeech.2019-1437}
}

@INPROCEEDINGS{ivector_asr2014,  author={Gupta, Vishwa and Kenny, Patrick and Ouellet, Pierre and Stafylakis, Themos},  booktitle={Proc. ICASSP},   title={I-vector-based speaker adaptation of deep neural networks for French broadcast audio transcription},   year={2014},  volume={},  number={},  pages={6334-6338},  doi={10.1109/ICASSP.2014.6854823}}

@ARTICLE{pcen,  author={Lostanlen, Vincent and Salamon, Justin and Cartwright, Mark and McFee, Brian and Farnsworth, Andrew and Kelling, Steve and Bello, Juan Pablo},  journal={IEEE Signal Processing Letters},   title={Per-Channel Energy Normalization: Why and How},   year={2019},  volume={26},  number={1},  pages={39-43},  doi={10.1109/LSP.2018.2878620}}

@ARTICLE{multichannel_asr2014,  author={Swietojanski, Pawel and Ghoshal, Arnab and Renals, Steve},  journal={IEEE Signal Processing Letters},   title={Convolutional Neural Networks for Distant Speech Recognition},   year={2014},  volume={21},  number={9},  pages={1120-1124},  doi={10.1109/LSP.2014.2325781}}

@INPROCEEDINGS{librispeech,  author={Panayotov, Vassil and Chen, Guoguo and Povey, Daniel and Khudanpur, Sanjeev},  booktitle={Proc. ICASSP},   title={Librispeech: An ASR corpus based on public domain audio books},   year={2015},  volume={},  number={},  pages={5206-5210},  doi={10.1109/ICASSP.2015.7178964}}

@inproceedings{ecapa_tdnn2020,
  author={Brecht Desplanques and Jenthe Thienpondt and Kris Demuynck},
  title={{ECAPA-TDNN: Emphasized Channel Attention, Propagation and Aggregation in TDNN Based Speaker Verification}},
  year=2020,
  booktitle={Proc. Interspeech},
  pages={3830--3834},
  doi={10.21437/Interspeech.2020-2650}
}

@inproceedings{rawnet2,
  author={Jee-weon Jung and Seung-bin Kim and Hye-jin Shim and Ju-ho Kim and Ha-Jin Yu},
  title={{Improved RawNet with Feature Map Scaling for Text-Independent Speaker Verification Using Raw Waveforms}},
  year=2020,
  booktitle={Proc. Interspeech},
  pages={1496--1500},
  doi={10.21437/Interspeech.2020-1011}
}

@INPROCEEDINGS{aasist2022,
  author={Jung, Jee-weon and Heo, Hee-Soo and Tak, Hemlata and Shim, Hye-jin and Chung, Joon Son and Lee, Bong-Jin and Yu, Ha-Jin and Evans, Nicholas},
  booktitle={Proc. ICASSP}, 
  title={{AASIST}: Audio Anti-Spoofing Using Integrated Spectro-Temporal Graph Attention Networks}, 
  year={2022},
  volume={},
  number={},
  pages={6367-6371},
  doi={10.1109/ICASSP43922.2022.9747766}}

@inproceedings{sahidullah16_interspeech,
  author={Md. Sahidullah and Héctor Delgado and Massimiliano Todisco and Hong Yu and Tomi Kinnunen and Nicholas Evans and Zheng-Hua Tan},
  title={{Integrated Spoofing Countermeasures and Automatic Speaker Verification: An Evaluation on ASVspoof 2015}},
  year=2016,
  booktitle={Proc. Interspeech},
  pages={1700--1704},
  doi={10.21437/Interspeech.2016-1280}
}
}

\end{document}